# Staged Evolution with Quality Gates for Model Libraries

Alexander Roth
RWTH Aachen University
Ahornstr. 55
52074 Aachen, Germany
roth@se.rwth-aachen.de

Andreas Ganser
RWTH Aachen University
Ahornstr. 55
52074 Aachen, Germany
ganser@swc.rwth-aachen.de

Horst Lichter
RWTH Aachen University
Ahornstr. 55
52074 Aachen, Germany
lichter@swc.rwth-aachen.de

Bernhard Rumpe
RWTH Aachen University
Ahornstr. 55
52074 Aachen, Germany
rumpe@se.rwth-aachen.de

## ABSTRACT

Model evolution is widely considered as a subject under research. Despite its role in research, common purpose concepts, approaches, solutions, and methodologies are missing. Limiting the scope to model libraries makes model evolution and related quality concerns manageable, as we show below.

In this paper, we put forward our quality staged model evolution theory for model libraries. It is founded on evolution graphs, which offer a structure for model evolution in model libraries through evolution steps. These evolution steps eventually form a sequence, which can be partitioned into stages by quality gates. Each quality gate is defined by a lightweight quality model and respective characteristics fostering reusability.

## Categories and Subject Descriptors

D.2.13 [**Software Engineering**]: Reusable Software—*Domain engineering, Reusable libraries, Reuse models*

## 1. INTRODUCTION

Omnipresent computers have lead to increasing complexity of programs and their development. In order to stay on top of things, numerous approaches were proposed and one among them is Model-Driven Development (MDD). It enables designing software by means of models, which are the results of mapping objects from the real world to a higher level of abstraction [25].

These models are often formulated in UML for two reasons. First, because "for larger and distributed projects, UML modeling is believed to contribute to shared understanding of the system and more effective communication" [5]. Second, UML models frequently serve as inputs for code generation, which helps prototyping, speeds up development, and supports generation of parts of the final system. This reduces the risks of working at an inappropriate level of abstraction or to get lost in details.

Since models are such essential assets in MDD, it is widely proposed to reuse models. This would decrease development time and increase software quality [22, 16]. Consequently, models should be stored in model libraries, which enable model management and allow for model reuse. A challenge regarding model libraries is to provide models of high quality, since model quality is a matter of subjectivity and hard to measure precisely [24]. Additionally, models in libraries undergo changes over time. Altogether this puts an extra burden on maintaining reusable models at high quality, which we investigate subsequently.

We believe that evolution and quality in model libraries is insufficiently supported and, to the best of our knowledge, we could not find an approach enabling guided model evolution. This is surprising, since model quality and evolution have been a subject under research for a long time already (sec. 2). Most importantly, model evolution is generally understood as a goal that is achieved only if the model has been adapted guaranteeing the system's integrity (sec. 2). In contrast, we believe that the understanding of quality and evolution needs to shift, when considered for model libraries (sec. 3). But this requires a more formal approach that supports evolution for model libraries (sec. 3.4) and helps to partition model states into stages according to their reusability (sec. 3.5). In this approach, each stage is separated by a gate, which is derived from a quality model and respective characteristics focusing on reusability (sec. 4).

All in all, we contribute a stage model enhanced with quality gates, which are defined by aspects of a quality model.

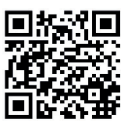

## 2. RELATED WORK

Model evolution, as we will present, can be discussed closely related to model libraries and quality of models. In the following, we present the current understanding of model evolution, model repositories, and model quality. However, we intend to look at model evolution in model libraries and will point out differences to the current understanding of model evolution.

Evolution of models is often considered as a goal one is trying to achieve automatically. A tool providing support for that is MoDisco [6] (cf. AM3 [1]), which aims at supporting evolution of legacy systems by model-driven approaches. That means, MoDisco is for re-engineering legacy software by means of models and starting a model driven development from gleaned models. In addition, issues of co-evolution are discussed. However, we keep to plain model evolution and propose that it is aimless and cannot be achieved. Consequently, it has to be guided in the right direction.

Another tool providing support for model evolution and co-evolution is COPE. It monitors evolution in an operation-based manner and applies traced changes to other models [12]. Moreover, these traces can be stored in a library so they can be forwarded or applied to other models as well. This is important as co-evolution is the focus of this project. Our approach differs slightly, since we do not need to monitor changes in the long run and focus on quality aspects and stages that depend on the actions performed on the models. But, some similarities between COPE and our approach can be found in regard of the stages underlying the approach. In the end, we need more formally defined stages, while COPE needs more formally defined operations. These operations are derived from oodb-schema experience and object-oriented source code experience, i.e., refactoring, and are overview over high-level operations or others. All together, operations are seen as structural or non-structural primitives. For example, specialization, generalization, delegation, or inheritance are such primitive operations. Others are replacement and merge-split [13].

Since we study evolution in model libraries, we need to discuss model management. Such systems offer mechanism to query for models, resolve conflicts, and manage versioning. As a consequence, aspects of co-evolution are usually out of scope. For example, MOOGLE is a model search engine with advanced querying functionality and user-friendly user interface support [20]. Second, ReMoDD aims at documentary purpose by offering models to the modeling community [9]. Third, the "Model Repository" manages models in a graph structure, holding, e.g., classes as vertices and associations as edges [28]. Fourth, AMOR is a model library providing advanced conflict resolution and versioning support [2]. None of these model libraries take model evolution into consideration. Consequently, we have implemented our own model library, which offers model evolution support.

But evolution support, compared to versioning, makes sense only if a common understanding of model quality is maintained. In conceptual modeling, quality is seen as a subjective and rather social than a formally definable property lacking a common understanding and standards [24]. Moody explains that the main challenge in defining model quality is the absence of measurable dimensions. Fortunately, model libraries limit the scope noticeably, so we can present an approach to measure model quality in model libraries. It is based on measurable quality dimensions and backed up by model metrics and reviews.

Supporting quality dimensions, we employ the results by Lindland et al. [19]. They define *syntactic*, *semantic*, and *pragmatic quality* as the cornerstones of their quality dimensions. Bansiya et al. describe that all of these quality dimensions can influence each other [3]. We extended Lindland's approach to take into account emotional quality.

Multiple approaches exist to define UML model quality. The most suitable for us is by Lange et al. [17]. They describe model quality as a set of dimensions and identified two primary use cases of models: *maintenance* and *development*. For each of these, model characteristics, which can influence multiple model purposes, are derived from the proposed set of dimensions. The characteristics have to be measured to draw conclusions about the model's quality. However, Mohagheghi et al. point out that measuring the presented characteristics is challenging [23]. These challenges are reduced when model evolution is considered for model libraries, since model quality can be limited to general quality.

## 3. MODEL EVOLUTION

Evolution has been a subject under research for decades. In contrast to its roots in biology, evolution theories are rather new to computer science. While Lehman summarized observations on software evolution in the eighties [18], it is questionable if these observations hold true for state-of-the-art software development. For example, Model Driven Development (MDD) changed software development in a way that some assumptions might need to be reconsidered. Consequently, we investigated interdisciplinarily and analyzed potential influences, bearing in mind that our area is model evolution in model libraries. We did so, to define this evolution, its characteristics, and the underlying formalism.

### 3.1 Influences of Model Libraries

Existing research in model evolution assumes that model evolution happens within software systems and addresses basically two concerns. On the one hand, collaboration is seen as one cause initiating model evolution. This is due to different developers and modelers working on the same model without a shared repository. As a consequence, these models are less frequently reviewed leading to declining model quality. On the other hand, integrity is seen to hamper model evolution, since adapted models need to preserve certain constraints. One of these constraints might be that system's integrity has to be preserved. In conclusion, these two concerns define model evolution as a goal.

Looking at model evolution in a context of model libraries and model reuse, the properties of model evolution change. This is due to models examined by different modelers disregarding any environment and purely focusing on reusability. Hence, the models are meant to provide solutions for problems preserving experience while ignoring any other integrity. As a result, the models are supposed to not contain technical details and maintain high quality. (cf. figure 1 left)

### 3.2 Lehman Laws in Model Evolution

The Lehman laws [18] offer four aspects which can be adapted to model evolution. First, model evolution is triggered, if a model needs to be changed. This might be due to changing requirements or bugs. Second, if no attempts are undertaken to decrease model complexity, it will be less often reused. The reason for that is mostly because the purpose of the model is hazed. Third, model evolution is feedback regulated; hence, evolution happens after the model has been reviewed and deployed. Last, except for adaptions that are necessary keep the mapping between a particular range of the real world and the model itself, the quality of a model is not influenced by a constant need for adaption. For instance, models that provide a suitable solution might not be changed but just reused.

As a consequence, models in a model library should focus on problem solving omitting technical aspects. In this regard, the size of a model is of vital importance for its reusability. We basically consider the size of a model as the number of classes, attributes, and methods. To that end, large models are harder to reuse than small models. Moreover, models that changed over time require reviews by other modelers to maintain reusability. Finally, model evolution does not inherit *hidden evolution*, i.e., changes in size always impact the systems visibly [29].

### 3.3 Model Evolution in Model Libraries

Since we limit model evolution to model libraries, we limit its start to the moment the model is added to a model library. Consequently, Lehman's definition of evolution as "a progressive and beneficial process of changes in evolving attributes of an entity" [18] needs refinement. This means, progressive changes need to be considered as *add*, *delete*, *rename*, and *retype* operations with the semantics defined in [12]. This correlates with Keienburg et al., who define evolution as "proceeding an ordered list of model change primitives on an existing component model and finally create a new model version" [15]. Adapting this to model evolution means that model evolution needs a definition of model versions, we call *model snapshots*. Each time a model is changed in a model library, the model has evolved and has formed a new snapshot. We define two consecutive model snapshots of the same model as an *evolution step*. A sequence of model snapshots represents *model evolution*, in our terms.

An example for model evolution is depicted in figure 1. It shows a class diagram modeling a jukebox. Starting in snapshot 1 the model is changed by applying rename, retype, delete, or add operations. This results in a snapshot 2. Further changes then might lead to several more snapshots and eventually the model will be in snapshot n. In this example, some technical details, as the suffix `DAO` indicates, spoiled snapshot 1 and should be removed first.

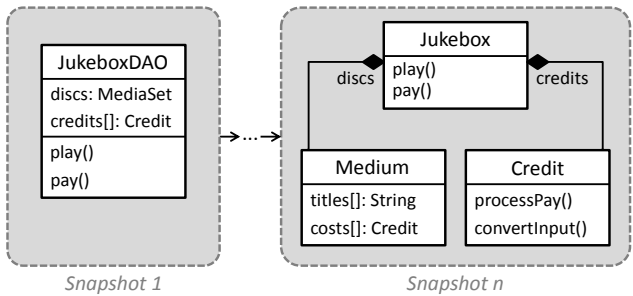

**Figure 1: Model Evolution Example**

It is important to understand that model evolution is not solely based on adaptations and it is to be distinguished from maintenance, because maintenance focuses on correcting faults, adapting, improving a model, and preserving integrity between models and source code [14]. Still, adaption is one major influence on model evolution, even if model evolution is understood as undirected, infinite, and unpredictable [4, 7, 21]. Taking a closer look at this kind of evolution, Judson et al. define three types [14]. Adaptive evolution happens due to changes in requirements, and perfective evolution comprises enhancements in the models design quality; corrective evolution summarizes error corrections.

These model evolution types characterize the reasons for evolution rather than describing activities as in maintenance. It is important to understand that model evolution is not maintenance, since maintenance changes models and thereby influences model evolution. In addition, maintenance is not the only reason for model evolution. A model can evolve due to structural changes that are forced by, e.g., successive development and not sustainment.

Another important aspect is that model evolution rarely happens without conflicts. Mens et al. subdivide these conflicts into two types: evolutionary and composition conflicts [22]. Evolutionary conflicts are conflicts describing the change of the model's behavior. This means, a property that holds in a model snapshot does not hold for its successor. Composition conflicts result from co-evolution of referenced models. In other words, if a model holds references to another model, these references need to be verified.

### 3.4 Model Evolution Graph

A first step in formalizing model evolution in model libraries is through structuring its progress. This can be achieved by adapting version graphs known from version control systems to model evolution. Eventually, this suits as the foundation for tool support and a methodology.

Derived from version control graphs, a *model evolution graph* represents a sequence of pairwise connected model snapshots. Each snapshot is contained in a *vertex* representing a model at a certain point in time as defined above. These vertices are labeled with unique *version ids* as known from version control systems. Moreover, each vertex has at least two adjacent edges, which are labeled with *transition ids*. A simple excerpt of a model evolution graph is depicted in figure 1. It shows that a model starts in a snapshot labeled with a version id `1` and progresses to snapshot `n`. Usually, each transition is labeled as well, e.g., with $\sigma_1$.

Each edge in a model evolution graph represents a sequence of primitive operations applied to a model while progressing to the next snapshot. In more detail, these primitive operations are defined as $\tau_{\text{add}}, \tau_{\text{delete}}, \tau_{\text{rename}}$, and $\tau_{\text{retype}}$. They form a sequence of primitive operations through $\sigma_i$ being a word defined as $\sigma_i \in \{\tau_{\text{add}}, \tau_{\text{delete}}, \tau_{\text{rename}}, \tau_{\text{retype}}\}^*$, where $i \in \mathbb{N}$.

Now, formally defining the entire model evolution graph, it comprises a tuple $(EG_{model}, s, t)$, where $EG_{model} = (V, E)$ is a graph with a vertex set $V$ and an edge set $E$ and $s, t$ are labeling functions. $s : V \to S$ is a function for some domain $S$ labeling each vertex. Each edge is labeled by the function $t : E \to T$, where $T = \{\tau_{\text{add}}, \tau_{\text{delete}}, \tau_{\text{rename}}, \tau_{\text{retype}}\}^*$. Altogether, this enables a definition of model evolution in model libraries, which we define as a sequence of evolution steps, which, in turn, are transitions in our model evolution graph. This means, model evolution in model libraries is a sequence of model snapshots; i.e., the evolution in figure 1 is the sequence $(1, ..., n)$.

## 3.5 Model Evolution Stages

We conduced a small field study with an exemplary model evolution, and participants said that model evolution graphs should contain status information for each snapshot. Consequently, a model evolution graph can be partitioned into sequences of evolution steps with equal status information. For example, the excerpt from figure 1 could show how a model evolves by going through the snapshot 1 with status "vague" and being in snapshot n with status "decent".

Rooting qualitative evaluations on "traffic lights" seemed most reasonable to our participants. As a result, the number of stages was chosen to be three, as it is intuitively obvious what a model in "red", "yellow" or "green" condition means in terms of reusability. As a nice side effect, the cognitive load of these three stages is negligible, since the number is small and the semantics of the colors is obvious. We choose the names to be *vague*, *decent*, and *fine*.

A vaguely reusable model is indetermined with respect to it's reusability and quality. Its label is "red", i.e., the stage is vague. This means, the model stays in the library for further processing and is offered for reuse, but the modeler needs to be cautious using this model. It might be that this model holds some specific suffixes (cf. "DAO" in figure 1), technology dependent elements, adapters for legacy use, or even errors. Despite all this uncleanliness, it was chosen to be put in the library for reuse since it is thought to be reusable in general. Due to these characteristics of "vague", every model put in the library starts in this stage.

As the major issues are fixed, the model progresses to be a decently reusable model. Its label is "yellow" now, i.e., the stage is decent. This means, the model does not hold any specifics or errors and is reusable in general. Still, it might be, that certain care needs to be taken if a modeler reuses this model, for example, the purpose does not perfectly match the intuition or a design decision is not clever. Even the layout might not be very appealing or the qualitative statements rely on assessments only but not on actual experience while reusing this model.

A finely reusable model is considered almost perfectly reusable and holds a "green" label indicating that the stage is fine. Now, the purpose is in line with the model and the quality is most reasonable. Notwithstanding that, models most likely will not be reusably "out of the box". This might be due to a template mechanism that requires a modeler to fill in the optional parts or due to adaption that is required by this particular reuse. If any operation other than renaming or retyping is applied, the model looses it's fine status.

## 3.6 Model Evolution Automaton

Taking into account the above mentioned reasoning and operations, the formal model of the model evolution stages can be transformed into an automaton as depicted in figure 2. It shows, how a model is initially put in a vague stage and that certain operations ($\tau_i$) can be performed until an identity operation ($id_M$) moves the model to decent stage and so forth. The considered operations comprise adding, deleting, renaming and retyping elements in the model. Their impact on the model is quite different, which limits the operations available on a model in fine stage. This is, because adding or deleting elements in a model in fine stage could alter the model entirely.

More formally, the automaton ($Q$, $\sum$, $Z$, $\delta$, $q_0$, $F$) comprises the following elements: $Q$ is the set of vertices and defined as $Q := \{\text{VAGUE}, \text{DECENT}, \text{FINE}\}$. $\sum$ is the collection of input symbols defined as $\sum := Q \times (T \cup \{id_M\})$ with $T := \{\tau_{\text{add}}, \tau_{\text{delete}}, \tau_{\text{rename}}, \tau_{\text{retype}}\}$. Moreover, $q_0$ is the defined starting state ($q_0 := \text{VAGUE}$) and $F$ is the empty set ($F := \emptyset$) since evolution is not regarded finite. Finally, $\delta$ is the transformation and defined as $\delta := \delta_v \cup \delta_d \cup \delta_f$, with

$$\delta_v := \{(\text{VAGUE}, \tau_i) \mapsto \text{VAGUE} \mid \tau_i \in T\}$$
$$\cup \{(\text{VAGUE}, id_M) \mapsto \text{DECENT}\},$$
$$\delta_d := \{(\text{DECENT}, \tau_j) \mapsto \text{DECENT} \mid \tau_j \in T\}$$
$$\cup \{(\text{DECENT}, id_M) \mapsto \text{VAGUE}\}$$
$$\cup \{(\text{DECENT}, id_M) \mapsto \text{FINE}\}, \text{ and}$$
$$\delta_f := \{(\text{FINE}, \tau_k) \mapsto \text{FINE} \mid \tau_k \in \{\tau_{\text{rename}}, \tau_{\text{retype}}\}\}$$
$$\cup \{(\text{FINE}, id_M) \mapsto \text{DECENT}\}$$
$$\cup \{(\text{FINE}, id_M) \mapsto \text{VAGUE}\}$$

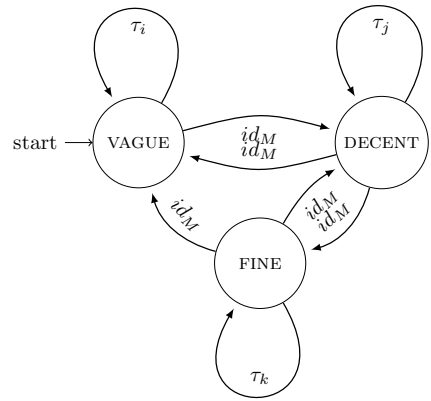

**Figure 2: Model Evolution Stages**

A closer look at the transitions unveils that some non-determinism is inherent in this automaton. First, transitions that go from the fine stage to another stage could represent the same type of operations. Second, the identity operations are no hindrance between stages. This means, so far it is always possible to change the stage and this is why user interaction based on quality assessments will be required later. But first, an example should provide a better idea how a model might progress through different stages.

## 3.7 Model Evolution in Action

The above mentioned automaton is non-deterministic; so, room is left for reasoning how this could be changed. We use the example from figure 1 and demonstrate how it might undergo evolution in a model library. This will set the reasoning required for transition guards between stages and eventually demand a quality model for decision support.

As our `jukebox` from figure 1 was considered beneficial for our model library, it was extracted from its environment and stored for reuse. Since most of the newly arriving models are expected to be biased in one way or another, the initial stage of our jukebox model is *vague*. In our example this is very reasonable, because the model contains a suffix that indicates its purpose and spoils the cleanness of this model. Without going into detail with quality in modeling, we can say that this model can be improved.

As a modeler improved the model to a certain extent, it can progress to "decent" stage. In our example, a modeler needs to remove the suffix `DAO` to make the model hold one class with a clean name. This was achieved by a simple rename operation and now the model is free from technological details. At this point, the modeler only needs to add, e.g., a description and model can progress to "decent" stage. This means, it is all right but certainly not "fine". Hence, the design should be revised, though this model is reusable in general. As a consequence, it is highly unlikely that it will be reused as it is. On the contrary, a modeler will most likely need to apply further add, delete, rename, or retype operations, if the model is reused. For example, some dependent classes, which obviously belong to this model, will need to be designed. In our example, a `Jukebox` is useless without a definition of a `MediaSet`.

As soon as the model matured further, it can progress to "fine" stage. This will have needed that some modelers reviewed this model or some of them removing model issues. Eventually, the gained snapshot might be like snapshot $n$ from figure 1. Mind that, this happened only with our default set of operations, which is mentioned above. The point is, that this model came to a stage that is "fine" for reuse. From now on, this model is valid in itself and can be reused by other modelers. Additionally, it might offer template support, which includes further modeling advise.

Since the quality of models and, likewise, template information breaks easily, when, e.g., add or delete actions are applied, only rename and retype actions are allowed in "fine" staged models. This is due to add and delete operations altering the model in a way that might totally alter the purpose and validity of that model. This is why these actions lower the stage immediately. But it depends on the reasoning of a modeler to decide on the next stage. Most likely, the model will only drop to a vague stage if it was split in two distinct parts, which need to be considered as "new" in the model library. Otherwise, add and delete operations will lead to "decent" stage.

From the above, we understand that some of the transitions need a conscious decision whether the model progresses to another stage. In order to help modelers to take these decisions, a quality model would be beneficial. But it is challenging to find such a model, since there is no precise definition of reusability for model libraries, nor is there a common understanding of what a quality model for model libraries might be. This is why we investigated further and found that this particular environment focusing on reusability limits certain aspects extensively, so that we were able to propose a reasonable solution.

## 4. QUALITY MODEL FOR MODELS

Restricting models to evolve in model libraries, where only general characteristics such as model size, i.e., complexity, are of relevance, model quality becomes manageable. If a model conforms to these general quality characteristics, the model is considered to have a high quality in a model library and is reusable. In the following, we present our quality model focusing on model libraries and thereby resolving the challenge of being hard to define and measure due to its highly subjective and relative nature [24].

Our quality model is based on the research of Lindland et al. [19] and Lange et al. [16] and comprising four general quality dimensions, namely *syntactic, semantic, pragmatic,* and *emotional quality*; each of which can be subdivided as shown in figure 3.

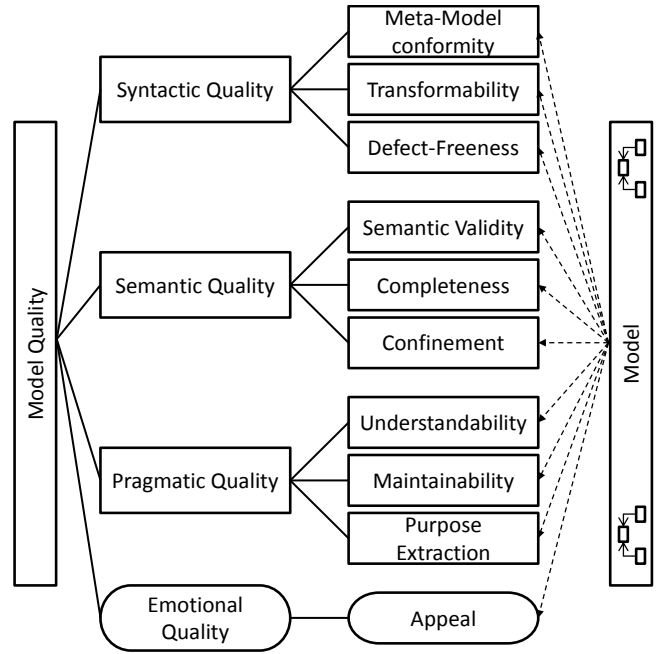

**Figure 3: Model Quality Dimensions**

First, focusing on reusability, it is sufficient to restrict syntactic quality to be described by the following three characteristics: *Meta-model conformity* describes the conformness of the model to the abstract syntax specifying the modeling language, i.e., the model is well-formed. Additionally, *transformability* states that a generator, which understands the model's syntax, can successfully transform a model into another representation, e.g., source code. Finally, a model is *defect-free*, if it does not contain any syntax errors. Moreover, syntactic quality can be used for error prevention, and error detection. Error correction is, however, not feasible, due to uncertain modeler goals.

Second, the extent to which the model describes the aspect it should model, i.e., correspondence between a model and its domain, is addressed by semantic quality. We subdivide this dimension into: *validity, completeness,* and *confinement*. A model is of high validity if, and only if, every statement defined by the model is valid in the domain. Whenever a model includes an invalid statement from the domain, it must be true that the advantage to simply remove this statement is higher than the drawbacks of adapting the whole model to the invalid statements. Furthermore, a model is complete if, and only if, there is no statement in the domain not yet included in the model, such that its benefit is higher than the drawback of adapting the model to this new statement. Third, if a model includes enough valid statements about the domain to describe the intention or to solve a problem in a domain, the model is confine.

Third, pragmatic quality addresses the correspondence between a modeler's comprehension and a model. In other words, the model must be understood by a modeler correctly. In model libraries, model comprehension can be reduced to three characteristics: *understandability, maintainability,* and *purpose extraction*. Understandability refers to the degree to which the modeler's interpretation can be build from the model. Maintainability addresses the degree to which a modeler understood the model and is able to adapt it to environmental changes. Finally, if the modeler's extracted purpose of a model corresponds to the purpose originally defined, the purpose extraction is high.

Last, emotional quality addresses the correspondence between a user's interpretation and his emotions. If a model is appealing to modelers, it is more likely to be reused. Consequently, such models have a high emotional quality. This model quality dimension is of importance in model libraries, since reuse depends on individual opinions about models. This model quality dimension is mentioned for the sake of completeness (cf. "thumb up button" in figure 5)

Each model quality dimension and especially the corresponding characteristics have to be measureable in order to enable qualitative statements about model quality in model libraries. SDMetrics [26] and Genero et al. [11] provide metrics to measure defect-freeness, understandability, and maintainability. Moreover, simplified model reviews can be used to measure semantic validity, completeness, confinement, and purpose extraction. Finally, meta-model conformity and transformability can be measured with existing validators [27]. Note that appeal is not measured but tracked with a counter, which counts the number of model reuses.

### 4.1 Quality Gates

The staged model evolution aims at structuring and guiding model evolution in model libraries. The structuring is achieved by partitioning model evolution in three stages. The traversal of models through the stages highly depends on the modeler and his intention. However, stage traversal can be guided by introducing *quality gates* in order to make qualitative statements about model quality in each stage and thereby derive its reusability. In the following, we summarize the staged model evolution approach as presented in section 3.5 and derive quality gates for the stage transition.

Generally, quality gates can be seen as metaphorical gates. A model can pass a gate, if all requirements of this gate are fulfilled. Figure 4 illustrates the staged model and the corresponding quality gates. The set of requirements for each gate is derived from the quality model for model libraries.

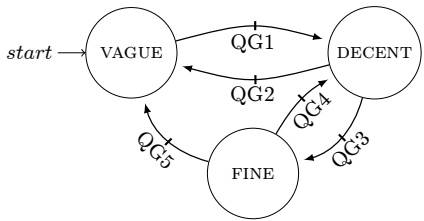

**Figure 4: Staged Model and Quality Gates**

Each model underlying the staged model evolution starts in the vague stage and is aiming at reaching the fine stage to be recommended for reuse in the model library. The main goal of the vague stage is to remove all obstacles resulting from the fact that the model has been added to the model library or has been restructured from a previously fine model. When all of these activities are finished and the modeler intends to move the model to the decent stage, the model has to pass the first quality gate QG1. The requirements of this quality gate are meta-model conformity, transformability, defect-freeness, semantical validity, and confinement. These requirements verify that (a) no prefixes and technological dependencies are included, (b) adapters for legacy are removed, and (c) the model has no errors. A model does not have to be semantically complete after passing this quality gate, because completeness is aimed for in decent stage.

When a model enters the decent stage, the staged evolution can be directed in two directions. First, if the modeler adapts the model to fit the initial purpose and is satisfied with the chosen layout and solution, the model can be moved to the fine stage by passing quality gate QG3. Since the next level of reusability is "green", i.e., the model library may suggest this model to modelers for reuse, this quality gate requires all quality characteristics to be fulfilled. Emotional quality, however, is not considered to be a requirement regarding the number of "likes", instead it shows that modelers agreed on the reuse of the model and its quality. Second, a model can be moved back to the vague stage due to errors in the model, restructuring, or even unsuitability of the model for reuse by passing quality gate QG2, which requires that at least one of the requirements for QG1 is violated.

After a model has been labeled to be reusable by entering the fine stage, different influences might lead to the loss of its status. If the model requires adaption other than renaming or retyping, it has to be moved to decent or vague stage depending on the type of adaption. Restructurings of the model, which lead to model decomposition or even new models, require the model to be moved to the vague stage. This also holds true for bugs in the model, which are hard to fix or are not going to be fixed in the future. In any other case of adaption, the model has to be moved to decent stage. In both cases, the model has to pass either quality gate QG5 or QG4. Both require that at least one of the requiremnts for QG3 is violated. This non-determinism is rooted in the adaptions required to be made to the model. E.g., if a model needs to be completely restructured, it is moved to vague stage.

## 5. PROTOTYPE

Based on the staged model evolution, a software prototype in Eclipse has been implemented and evaluated. The main focus of the prototype is guidance of the modeler through the staged model evolution. Hence, a simplified representation of the staged model (cf. figure 4) was implemented and enhanced with some guidance to make the quality model work. For example, model metrics and model reviews are included to provide measures regarding the quality model introduced above. In addition, wizards check and display the status of each quality gate, whenever the modeler intends to change the current stage. An overview of the main elements of the prototype is illustrated in figure 5.

The simplified representation of the stages (top of the figure) contains a representation for each stage colored with respect to its reusability status, i.e., vague is red, decent is yellow, and fine is green. The current stage and the transitions a modeler can choose to reach another stage are highlighted. This color encoding helps displaying the limited actions a modeler can perform in each stage (cf. figure 2) and thereby increases guidance.

In addition to the three stages of the staged model evolution, it is always possible that a model can become deprecated. Then, the model is kept in its current stage and marked as deprected. The basket symbol in figure 5 allows the modeler to always mark the current model as deprecated. Please note that once a model becomes deprecated,

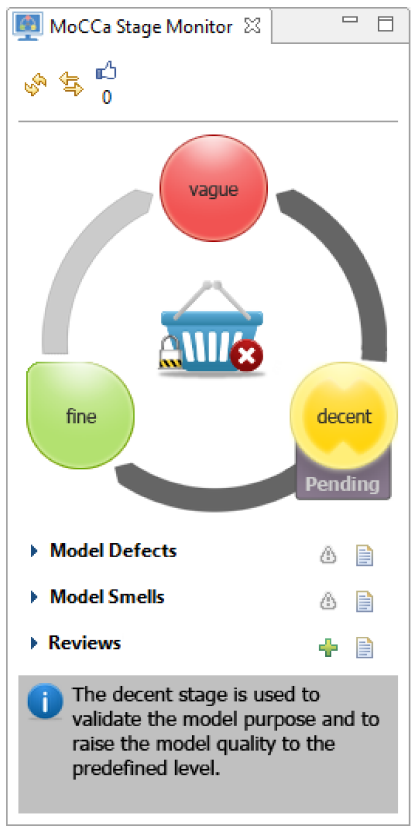

Figure 5: Staged Model Evolution Prototype

its stage cannot be changed anymore.

Besides the stage representation, the prototype contains a list of model metric values and model review reports to display information about the model quality. We separated model metrics in two categories: *model defects* and *model smells*. The first type of metrics indicates violations of metrics that are known to be errors, e.g., a new class is introduced without a class name. In contrast, model smells are defined based on Fowler's definition of smells [8]. But with respect to the modeling domain, a smell refers to an improvement of the model, e.g., reducing the amount of classes to make the model more understandable. Moreover, we simplified model reviews to be easy to create and to correspond to the model quality characteristics. At this point, we neglect presenting the mapping of a model review to a model quality characteristic, but point out the main idea of the simplifed model reviews. A simpified model review for each model quality characteristic is created by limiting the view of the review to the necessary elements required to verify the model characteristic.

The use of model metrics requires modelers to either manually apply them or an approach to automatically apply model metrics and inform the modeler about the results. A major drawback of a manual approach is that metrics will be forgotten over time and thus will not be applied. Consequently, we followed the approach of *proactive quality guidance*. This approach automatically calculates model metrics whenever a model has been changed and immediately displays metric violations, i.e., the modeler receives live feedback during modeling.

Another benefit of this approach is a set of clear instructions as the result of applying the metrics. These instructions inform the modeler about steps that need to be done to resolve the corresponding model metric violation. In combination with model reviews, which are required to be fulfilled in order to move to another stage, this proactive quality guidance approach provides non-obtrusive feedback to raise model quality awareness and helps guiding modelers through the staged model evolution.

## 6. CONCLUSION AND OUTLOOK

Despite a long history in model evolution research, there is no holistic approach available supporting evolution in model libraries. This is surprising, since models change in libraries as well. Moreover, it is surprising because these circumstances ease a few obstacles. Evolution, in general, is considered as aimless, but model libraries would loose their benefits quickly evolving unguided. Hence, evolution needs to be guided in model libraries, which offers a clear direction of evolution and provides opportunities to formalize and structure evolution in model libraries as we did above.

For structuring this evolution, we defined an evolution graph and partitioned it based on model stages. These stages serve as indicators for model reusability. We defined these stages to be vague, decent, and fine and attribute them with the colors red, yellow, and green respectively. This intuitive stage model allows to define gates that need to be fulfilled before passing to the next better stage.

In order to define quality gates to separate model stages, we introduced a quality model, which defines simple means of quality assurance. This is possible, because model libraries limit possible quality attributes and, hence, keep the scope of metrics small. All in all, we defined syntactic, semantic, and pragmatic quality and added emotional quality, which proved reasonable in our experiments.

All together, we provided a foundation for quality staged model evolution in model libraries, which we implemented in tool support providing guidance for this holistic approach.

Besides evaluation of this approach, a novel and altered kind of co-evolution is subject to current and future work. Consider a model library that is more a knowledge base. It would certainly hold information how single models are related to each other. For example, relationships could be established between such models on different levels.

We created such a model library [10] and enhanced it with evolution support as presented here. The drawback of this approach is that it only takes into account a single model disregarding any crossing relationships. But how do these relationships "co-evolve" as models evolve? Moreover, we analyze what happens if a model needs to split. This might happen, most likely, in fine stage and this is why the symbol in our tool, which stands for this stage, is shaped slightly different. As the modeler hovers over this symbol, a splitting arrow appears allowing to split this model in several parts. But the semantics and the consequences of what that would or could mean are still under research.

## 7. ACKNOWLEDGMENTS

We would like to thank all our anonymous reviewers for their comments and effort. Moreover, we would like to thank Andrej Dyck, Veit Hoffmann, and Matthias Vianden. It was always a great to have another intriguing discussion.